\documentclass[11pt]{article}
\usepackage{moriond,epsfig}

\def\be{\begin{equation}}
\def\ee{\end{equation}}
\def\bea{\begin{eqnarray}}
\def\eea{\end{eqnarray}}

\begin{document}
\vspace*{4cm}
\title{Two body hadronic $D$ decays}

\author{Yu Fusheng, Cai-Dian L\"{u}, Xiao-Xia Wang}

\address{Institute of High Energy Physics and Theoretical
Physics Center for Science Facilities, Chinese Academy of Sciences,
Beijing 100049, People's Republic of China}

\maketitle\abstracts{ We analyze the decay modes of $D/D_s\to PP,PV$
on the basis of a hybrid method with the generalized factorization
approach for emission diagrams and the pole dominance model for the
annihilation type contributions. Our results of PV final states are
better than the previous method, while the results of PP final
states are comparable with previous diagrammatic approach. }

\section{Introduction}

The CLEO-c and the two B factories already give more measurements of
charmed meson decays than ever. The BESIII and super B factories are
going to give even much more data soon. Therefore, it is a good
chance to further study the nonleptonic two-body $D$ decays.
However, it is theoretically unsatisfied since some model
calculations, such as QCD sum rules or Lattice QCD, are ultimate
tools but formidable tasks. In $B$ physics, there are QCD-inspired
approaches for hadronic decays, such as the perturbative QCD
approach (pQCD),\cite{pqcd} the QCD factorization approach
(QCDF),\cite{qcdf} and the soft-collinear effective theory
(SCET).\cite{scet} But it doesn't make much sense to apply these
approaches to charm decays, since the mass of charm quark, of order
1.5 GeV, is neither heavy enough for a sensible $1/m_c$ expansion,
 nor light enough for the application of chiral perturbation theory.

After decades of studies, the factorization approach is still an
effective way to investigate the hadronic $D$ decays \cite{BSW}.
However, the naive factorization encounters well-known problems: the
Wilson coefficients are renormalization scale and $\gamma_5$-scheme
dependent, and the color-suppressed processes are not well predicted
due to the smallness of $a_2$. The generalized factorization
approaches were proposed to solve these problems, considering the
significant nonfactorizable contributions in the effective Wilson
coefficients \cite{gf}. Besides, in the naive or generalized
factorization approaches,
 there are no strong phases between different amplitudes,
  which are demonstrated to be existing by experiments.

On the other hand, the hadronic picture description of non-leptonic
weak decays has a longer history, because of their non-perturbative
feature. Based on the idea of the vector dominance, which is
discussed on strange particle decays,\cite{vectordominance} the
pole-dominance model of two-body hadronic decays was
proposed.\cite{polemodel} This model has already been applied to the
two-body nonleptonic decays of charmed and bottom mesons
\cite{polemodel,Kramer:1997yh}.

In this work, the two-body hadronic charm decays are analyzed based
on a hybrid method with the generalized factorization approach for
emission diagrams and the pole dominance model for the annihilation
type contributions \cite{Fusheng:2011tw}.

\section{The hybrid method}

In charm decays, we start with the weak effective Hamiltonian for the $\Delta C=1$ transition
 \be\label{eq:Hami}
\mathcal{H}_{eff}=\frac{G_F}{\sqrt{2}}V_{CKM}(C_1O_1+C_2O_2)+h.c.,
 \ee
with the current-current operators
 \bea
O_1=\bar{u}_{\alpha}\gamma_{\mu}(1-\gamma_5)q_{2\beta}\cdot\bar{q}_{3\beta}\gamma^{\mu}(1-\gamma_5)c_{\alpha},
\nonumber\\
O_2=\bar{u}_{\alpha}\gamma_{\mu}(1-\gamma_5)q_{2\alpha}\cdot\bar{q}_{3\beta}\gamma^{\mu}(1-\gamma_5)c_{\beta}.
 \eea
In the generalized factorization method, the amplitudes are separated into two parts
 \be
\langle M_1M_2|\mathcal{H}_{eff}|D\rangle=\frac{G_F}{\sqrt{2}}V_{CKM}a_{1,2}\langle M_1|\bar{q}_1\gamma_{\mu}(1-\gamma_5)q_2|0\rangle
\langle M_2|\bar{q}_{3}\gamma^{\mu}(1-\gamma_5)c|D\rangle,
 \ee
where $a_1$ and $a_2$ correspond to the color-favored tree diagram
($\mathcal{T}$) and the color-suppressed diagram ($\mathcal{C}$)
respectively. To include the significant non-factorizable
contributions, we take $a_{1,2}$ as scale- and process-independent
parameters fitted from experimental data. Besides, a large relative
strong phase between $a_1$ and $a_2$ is demonstrated by experiments.
Theoretically, the existence of large phase is reasonable for the
importance of inelastic final state interactions in the charmed
meson decays, with  on-shell intermediate
 states. Therefore, we take
 \be
 a_1=|a_1|,~~a_2=|a_2|e^{i\delta},
 \ee
where $a_1$ is set to be real for convenience.

\begin{figure}[thpb]
 \begin{center}
\psfig{figure=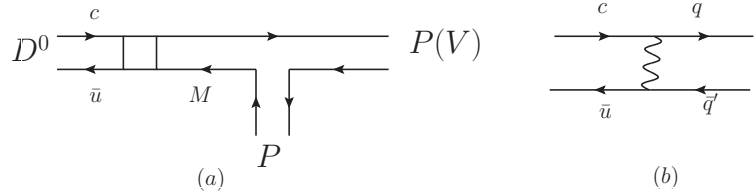,height=1 in}
\caption{Annihilation diagram in the pole-dominance model}
\label{fig:pole}
\end{center}
\end{figure}

On the other hand, annihilation type contributions are neglected in the factorization approach. However, the weak
annihilation ($W$-exchange and $W$-annihilation)
 contributions are sizable, of order $1/m_c$, and have to be considered. It is also demonstrated to be important by
 the difference of life time between $D^0$ and $D^+$.
 The pole-dominance model is a useful tool to calculate the considerable resonant effects of annihilation diagrams.
 For simplicity, only the lowest-lying pole is considered in the single-pole model. Taking $D^0\to PP,PV$ as example,
 the annihilation type diagram
 in the pole model is shown in Fig.\ref{fig:pole}(a). $D^0$ goes into the intermediate state $M$ via the effective
 weak Hamiltonian in Eq.(\ref{eq:Hami}),
 shown by the quark line in the Fig.\ref{fig:pole}(b), and then decays into $PP(PV)$ through strong interactions.
 Angular momentum should be conserved at the weak vertex,
 and all conservation laws be preserved at the strong vertex. Therefore, the intermediate particles are scalar mesons
  for $PP$ modes and pseudoscalar mesons for $PV$ modes.
In $D^0$ decays, they are $W$-exchange diagrams, but $W$-annihilation amplitudes in the $D^+_{(s)}$ decay modes.

The weak matrix elements are evaluated in the vacuum insertion approximation\cite{Kramer:1997yh},
 \bea
 \langle M|\mathcal{H}|D\rangle&=&\frac{G_F}{\sqrt{2}}V_{CKM}a_{A,E}\langle M|\bar{q}_1\gamma_{\mu}(1-\gamma_5)q_2|0\rangle
\langle 0|\bar{q}_{3}\gamma^{\mu}(1-\gamma_5)c|D\rangle
\nonumber\\
&=&\frac{G_F}{\sqrt{2}}V_{CKM}a_{A,E}f_M f_D m_D^2,
 \eea
where the effective coefficients $a_A$ and $a_E$ correspond to $W$-annihilation and $W$-exchange amplitudes respectively.
Strong phases relative to the emission diagrams are also considered in these coefficients.

For the $PV$ modes, the effective strong coupling constants are defined through the Lagrangian
\begin{equation}
\mathcal{L}_{VPP}=ig_{VPP}V^{\mu}(P_1{\partial}_{\mu}P_2-P_2{\partial}_{\mu}P_1),
\end{equation}
where $g_{VPP}$ is dimensionless and obtained from experiments. By inserting the propagator of the intermediate
 state $M$, the annihilation amplitudes are
\be \langle
PV|\mathcal{H}_{eff}|D\rangle=\frac{G_F}{\sqrt{2}}V_{CKM}a_{A,E}f_M
f_D m_D^2\frac{1}{m_D^2-m_M^2} g_{VPM} 2(\varepsilon^*\cdot p_D).
\ee
As for the $PP$ modes, the intermediate mesons are scalar
particles. The effective strong coupling constants are described by
\begin{equation}
\mathcal{L}_{SPP}=-g_{SPP}m_{S}SPP.
\end{equation}
However, the decay constants of scalar mesons are very small, which
is shown in the following relation
\begin{eqnarray}
\frac{f_S}{\bar{f}_S}=\frac{m_2(\mu)-m_1(\mu)}{m_S}, \label{fs2}
\end{eqnarray}
where $f_S$ is the vector decay constant used in the pole model, $\bar f_{S}$ is the scale-independent scalar decay
 constant, $m_{1,2}$ are the running current quark mass,
and $m_S$ is the mass of scalar meson. Therefore, the scalar pole contribution is very small, resulting in little
resonant effect of annihilation type contributions
in the $PP$ modes. On the contrary, large annihilation contributions are given in the $PV$ modes by relative large
 decay constants of intermediate pseudoscalar mesons.

\section{Numerical results and discussions}

In this method, only the effective Wilson coefficients with relative
strong phases are free parameters, which are chosen to obtain the
suitable results consistent with experimental data. For $PP$ modes,
\begin{eqnarray} \label{eq:1}
         a_1 &=& 0.94\pm0.10,~~~~~~~~~~~~~~~~~~
         a_2 = (0.65\pm0.10) e^{i(142\pm10)^{\circ} }, \nonumber\\
         a_A &=& (0.20\pm0.10) e^{i(300\pm10)^{\circ} },~~
         a_E = (1.7\pm0.1) e^{i(90\pm10)^{\circ} }.
\end{eqnarray}
For $PV$ modes,
\begin{eqnarray} \label{eq:2}
         a_1^{PV} &=& 1.32\pm0.10,~~~~~~~~~~~~~~~~~
         a_2^{PV} = (0.75\pm0.10) e^{i(160\pm10)^{\circ} }, \nonumber\\
         a_A^{PV} &=& (0.12\pm0.10) e^{i(345\pm10)^{\circ} },~~
         a_E^{PV} = (0.62\pm0.10) e^{i(238\pm10)^{\circ} }.
\end{eqnarray}

All the predictions of the 100 channels are shown in the tables of
ref.\cite{Fusheng:2011tw}. The prediction of branching ratio of the
pure annihilation process $D_s^+\to\pi^+\pi^0$ vanishes in the pole
model within the isospin symmetry. It is also zero in the
diagrammatic approach in the flavor SU(3) symmetry. Simply, two
pions can form an isospin 0,1,2 state, but 0 is ruled out
 because of charged final states, and isospin-2 is forbidden for the leading order $\Delta C=1$ weak decay.
 The only left s-wave isospin-1 sate is forbidden by Bose-Einstein statics. In the pole model language, $G$ parity is violated
 in the isospin-1 case. Therefore, no annihilation
 amplitude contributes to this mode.

The theoretical analysis in the $\eta-\eta'$ sector is kind of complicated. The predictions with $\eta'$ in the final
state are always smaller
in this hybrid method than those case of $\eta$
due to the smaller phase space. However, it is opposite by experiments in some modes, such as $D_s^+\to\pi^+\eta(\eta')$,
 $D^0\to\bar K^0\eta(\eta')$.
This may be the effects of SU(3) flavor symmetry breaking for
$\eta_q$ and $\eta_s$, the error mixing angle between $\eta$ and
$\eta'$ \footnote{The theoretical and phenomenological estimates for
the mixing angle $\phi$ is $42.2^\circ$ and $(39.3 \pm 1.0)^\circ$,
respectively.\cite{Feldmann:1998vh}}, inelastic final state
interaction, or the two gluon anomaly mostly associated to the
$\eta'$, etc.. The mode of $D_s^+\to\rho^+\eta(\eta')$ is similar
with the above two cases, the opposite ratio of $\eta$ over $\eta'$
between theoretical prediction and the data. But this is a puzzle by
experiment measurement, which is taken more than ten years ago
\cite{cleo}. As is questioned by PDG \cite{PDG2010}, this branching
ratio of $(12.5\pm2.2)\%$ considerably exceeds the recent inclusive
$\eta'$ fraction of $(11.7\pm1.8)\%$.

Recently, model independent diagrammatic approach is used  to
analyze the charm decays \cite{Rosner:1999xd}. All two-body hadronic
decays of $D$ mesons can be expressed in terms of some distinct
topological diagrams within the SU(3) flavor symmetry, by extracting
the topological amplitudes from the data \cite{Cheng:2010ry}. Since
the recent measurements of $D_s^+\to\pi^+\rho^0$ \cite{:2008tm} and
$D_s^+\to\pi^+\omega$ \cite{:2009vk} give a strong constraint on the
$W-$annihilation amplitudes, one cannot find a nice fit for $A_P$
and $A_V$ in the diagrammatic approach to the data with $D_s^+\to
\bar K^{*0}K^+,\bar K^0K^{*+}$ simultaneously. Compared to the
calculations in the model-independent diagrammatic approach
\cite{Cheng:2010ry}, our hybrid method
 gives more predictions for the $PV$ modes in which the
predictions are consistent with the experimental data. It is
questioned that the measurement of
 $Br(D_s^+\to\bar K^0K^{*+})=(5.4\pm1.2)\%$,\cite{Chen:1989tu} which was taken two decades ago, was
overestimated.
 Since $|C_V|<|C_P|$ and $A_V\approx
A_P$ as a consequence of very small rate of $D_s^+\to\pi^+\rho^0$,
it is expected that $Br(D_s^+\to\bar K^0K^{*+})<Br(D_s^+\to \bar
K^{*0}K^+)=(3.90\pm0.23)\%$. Our result in the hybrid
 method also agrees with this argument.

 As an application of the diagrammatic approach, the mixing parameters
$x=(m_1-m_2)/\Gamma$ and $y=(\Gamma_1-\Gamma_2)/\Gamma$ in the
$D^0-\bar D^0$ mixing are evaluated from the long distance
contributions of the $PP$ and $VP$ modes \cite{Cheng:2010rv}. The
global fit and predictions in the diagrammatic approach are done in
the SU(3) symmetry limit. However, as we know, the nonzero values of
$x$ and $y$ come from the SU(3) breaking effect. Part of the flavor
SU(3) breaking effects are considered in the factorization method
and in the pole model. Therefore, our hybrid method takes its
advantage in the analysis
 of $D^0-\bar D^0$ mixing.

\section*{Acknowledgments}

This work is partially supported by National  Science Foundation of
China under the Grant No. 10735080, 11075168; and National Basic
Research Program of China (973)
  No. 2010CB833000.

\end{document}